# Measuring the complexity of micro and nanostructured surfaces

A. Arapis[a], V. Constantoudis[a,b], D. Kontziampasis[c,d,e], A. Milionis[f], C.W.E. Lam[f], A. Tripathy[f], D. Poulikakos[f], E. Gogolides[a,b]

[a]*Institute of Nanoscience and Nanotechnology, NCSR Demokritos, Neapoleos 27, Agia Paraskevi, Greece*
[b]*Nanometrisis p.c., Neapoleos 27, Agia Paraskevi, Greece* [c]*School of Biomedical Sciences, Faculty of Biological Sciences, University of Leeds, LS29JT, Leeds, UK* [d]*Astbury Centre for Structural and Molecular Biology, University of Leeds, LS29JT, Leeds, UK* [e]*Department of Engineering, School of Digital, Technolgies and Arts, Staffordshire University, ST42EF, Stoke-on-Trend, UK* [f]*Laboratory of Thermodynamics in Emerging Technologies, Department of Mechanical and Process Engineering, ETH Zurich, Sonneggstrasse 3, 8092 Zurich, Switzerland*

Nanostructured surfaces usually exhibit complicated morphologies that cannot be described in terms of Euclidean geometry. Simultaneously, they do not constitute fully random noise fields to be characterized by simple stochastics and probability theory. In most cases, nanomorphologies consist of complicated mixtures of order and randomness, which should be described quantitatively if one aims to control their fabrication and properties. In this work, inspired by recent developments in complexity theory, we propose a method to measure nanomorphology complexity that is based on the deviation from the average symmetry of surfaces. We present the methodology for its calculation and the validation of its performance, using a series of synthetic surfaces where the proposed complexity measure obtains a maximum value at the most heterogeneous morphologies between the fully ordered and fully random cases. Additionally, we measure the complexity of experimental micro and nanostructured surfaces (polymeric and metallic), and demonstrate the usefulness of the proposed method in quantifying the impact of processing conditions on their morphologies. Finally, we hint on the relationship between the complexity measure and the functional properties of surfaces.

## 1. Introduction

Nanotechnology is largely based on our ability to pattern surfaces at the nanometer scale. A key challenge to control the nanopatterning process and justify its repeatability, is the quantitative characterization of the morphology of the produced nanostructured surfaces. However, to achieve this, we need to surmount the obstacle of understanding and controlling the complicated morphologies that nanopatterned surfaces usually exhibit. Even in the case of top-down techniques, the output patterns deviate from the initially designed well-defined shapes due to the presence of random-like shape undulations. On the other hand, bottom-up approaches usually result in stochastic morphologies that frequently exhibit self-organized and semi-ordered structures. These differ from the fully random scenario, where morphology resembles uncorrelated noise. Therefore, a typical situation in nanopatterning is when nanomorphologies consist of both order and randomness, leading to an extremely rich palette of surface architecture. Although this interplay between order and randomness seems to be inevitable and in some cases critical in underpinning desired functionalities, up to now there is no systematic and targeted methodology developed to quantify this.

The most common approaches for the characterization of stochastic aspects of nanomorphologies are based on roughness theory, or more rarely on fractal geometry. In roughness theory, the main focus is on the quantification of surface deviations from full flatness, while emphasis is given on the role of their amplitude (vertical) and spatial (horizontal) aspects. Nevertheless, no special attention is given on the characterization of the intricate co-existence of order and randomness. The most relevant parameters for such a characterization are the correlation length and the entropy. The first is calculated by means of the autocorrelation function and quantifies the range of surface correlations, while the more rarely used entropy of surface heights is estimated through their distribution function. Both parameters can be considered in order to characterize the degree of randomness in surface morphologies, since they decrease (correlation length) or increase (entropy) monotonically with randomness. For this reason, they do not meet the requirement of a dedicated metric for the characterization of the order-randomness co-existence. Such a metric should be maximized between full order and randomness, given that both share the same property of morphological homogeneity. Fig. 1 shows schematically the motivation behind the idea of seeking a metric that is maximized between order and randomness, at the point where morphology becomes more heterogeneous and unpredicted.

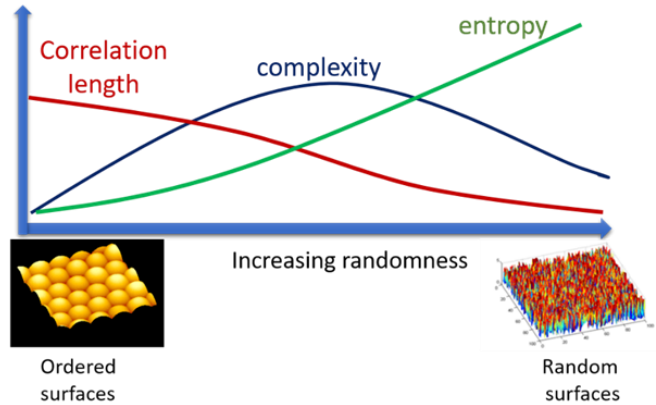

Fig. 1: The diagram shows schematically the desired behavior of the complexity measure which should exhibit a maximum between the extrema of full order and randomness (blue curve), contrary to the monotonic decrease of the correlation length (red curve) and the monotonic increase of entropy (green curve)

Fractal geometry assumes the existence of scaling symmetry in geometrical objects and elaborates a methodology for the characterization of an object structure, taking advantage of the properties of this symmetry. However, in nanomorphologies the scaling symmetry (when present) is usually restricted to a narrow range of scales. Therefore, though fractal approach provides an insightful look at structures with complicated geometries, its usefulness in the characterization of nanomorphologies is severely confined by its nature.

On the other hand, one may recognize that the demand for a metric characterizing structures between order and randomness is not restricted to nanometrology and the quantification of nanomorphologies. Actually, it lies at the heart of the complexity science, since the synergy of order and randomness is considered one of the hallmarks of complex systems and structures [3-6]. In order to find a measure of that synergy, recent developments bring forward the concept of average (or statistical) symmetry [7, 8]. The latter allows us to put on the same page ordered and random structures, since they both obey translational symmetry and become homogeneous when averaging over some scale. In this view, the complexity is defined as the deviation of a structure from the average symmetry and it is expected to be maximized in between the extremes of full order and randomness. Therefore, it seems to fit perfectly to the need of nanometrology for a metric that is specialized on the characterization of the complexity of nanomorphologies.

The key idea of the paper, and inspired by the aforementioned developments, is to propose a complexity-based approach for the characterization of complicated nanomorphologies through the analysis of their microscopy images. To this end, we first present the mathematical methodology for the estimation of the complexity measure, which is based on the notion of average symmetry, as well as the modifications we made to properly adapt it to the specific needs of the microscopy images of nanomorphologies. This is the subject of section 2. Subsequently, we validate the proposed methodology and metric using a series of synthesized surfaces, which are described and analyzed in section 3. Section 4 presents the first applications of this complexity-based approach, using it on laboratory fabricated nanostructured surfaces that are imaged either via Atomic Force Microscopy (AFM) or Scanning Electron microscopy (SEM). As a second example of experimental surfaces (micro and nanostructured Al surfaces), we link the complexity measure with their heat transfer coefficient.

## 2. Methodology

As described in the Introduction, in this work we propose a method that introduces a measure of quantification of the complexity of a system, using the deviation of its structure from the average or statistical symmetry at all scales. In particular, we follow the entropy-based implementation of this concept as it was presented in the Alamino's papers [7,8].

In the following section, we present Alamino's approach with a step by step description, since it results in the estimation of the complexity measure. We then focus on our methodology and the points that are differentiated from Alamino's implementation, in order to 'capture' the particularities of nanostructured surfaces as imaged in microscope images.

### 2.1 Alamino's approach: Average symmetry and entropy-based Complexity

Alamino presents the methodology for the definition and estimation of the complexity measure using the example of a N-dimensional grid with binary values on its nodes (black or white). For each node, a scale-dependent mass is calculated via a sphere of radius $r$, which defines the scale. The mass of the i-th node, denoted as $\mu_r(i)$, equals to the total number of the black nodes that are found on the surface of the sphere with radius $r$ and center the i-th node. In that manner, for every scale $r$, a mass field $\mu_r(i)$ is calculated. For computational reasons, the sphere can be replaced by a cube of side $r$.

When the mass fields at all scales $r$ are calculated, the normalized distributions can be estimated via the histograms of mass $m$

$$\lambda(m|r) = \frac{1}{N} * \sum_{i=1}^{N} \delta(m, \mu_r(i)) \quad (1)$$

, where $\delta$ is the Kroenecker delta, and $N$ are the total number of grid points where the analysis is conducted on.

Finally, the complexity $A(L, R)$ is calculated as the average entropy of the distributions that are derived for all system scales:

$$A(L, R) = -\frac{1}{R+1} * \sum_{r=0}^{R} \sum_m \lambda(m|r) * \ln \lambda(m|r) \quad (2)$$

where $L$ is the system size and $R$ is the maximum scale used in averaging. It is noted that the complexity depends on both the size of the system and the maximum scale that is considered.

### 2.2 Our methodology

Scanning Electron Microscopy (SEM) and Atomic Force Microscopy (AFM) are the most widespread methods that are used for the inspection and the measurement of the morphology of nanostructured surfaces. SEM is used to produce digital images from the signal of the emitted secondary electrons, and AFM scans a measurement area of a surface to take height measurements at the nodes of a 2D grid, and outputs the surface's topography. In accordance with Alamino's example, both microscopes output measurements at the nodes of a 2D grid, with resolution defined by the pixel size in SEM images, and by the sampling interval in AFM images. However, contrary to Alamino's grid, every point of the microscope measurements has values that are either multiple discrete, such as the pixel intensities of SEM images (ranging between 0 and 255), or continuous, such as the heights that are derived from the AFM topography measurements. Therefore, this methodology for the measurement of complexity, needs to be reformulated, taking into account the multiple discrete and continuous values of SEM and AFM microscope measurements respectively, if it is to be used for their analysis. A possible solution would be to convert the images to binary, by using a properly defined threshold. However, this conversion diminishes topographic information in the morphology measurements and therefore it is more preferable to modify Alamino's formulae, in order to consider multiple discrete or continuous values. Another solution could be to adapt a different complexity measure, such as the one presented by D. Zanette [9], which is relied on the analysis of local variations of the intensity of the pixels and the calculation of the average variance.

A second difference from Alamino's proposal refers to the boundary conditions that should be considered for the mass calculation, since they are based on scale averaging. Alamino tackled boundary conditions via periodicity. However, this approach is not easily and straightforwardly applied to the images of nanomorphologies since these are characterized by spatial correlations, which will be distorted if periodic conditions are applied. Special attention should be given on the continuity of image values. This is the reason behind why we prefer to apply the so-called border method [10] according to which we limit the range of analyzed area removing from the image area a zone around edges with width equal to the largest scale R included in the entropy averaging of eq.2.

For instance, a system which contains a grid of side $L$ and maximum scale $R$ the points $(x,y)$ for which the mass can be defined is:

$$(x, y) = \{(x, y) | \ x \in (R + 1: L - R), y \in (R + 1:)\} \quad (3)$$

The mass calculation step can be performed via at least three ways considering:

1. All the interior and all the side points of the square,
2. The side points of the square,
3. A selective fraction of the side points of the square,

Alamino chooses to work with the second approach, but in this work we adopt the third, since the consideration of all points at different scales constitute some form of bias when the scale increases. This is due to the fact that for the first and second approaches, the increase of the scale results in the increase of the points that are considered, and therefore, the calculation of the average is not only relied on the different scale, but also on the increase of the number of points that are considered. This could result in a complexity decrease due to the increase of points, while increasing the scale [11]. Hence, each mass is calculated as the average of 8 neighboring points (centers of the sides are found at $r$ distance, and corners at $\sqrt{2} * r$ distance) and itself. The mass field $m(x, y)$ for each scale $r$ is defined as follows:

$$I = \{x - r, x, x + r\} \quad (4)$$

$$J = \{y - r, y, y + r\} \quad (5)$$

$$m(x, y: r) = \frac{1}{9} \sum_{i \in I} \sum_{j \in J} z(i, j) \quad (6)$$

After the mass field calculation, mass distributions at all scales are obtained via the corresponding histograms. At this point, the last difference of the two approaches lies and refers to the binning. We calculate the mass values range at each scale and build bins of equal widths for all scales. The width of each bin depends on the application, and its choice is made in a way that can capture the necessary information regarding the morphology and/or topography of the surface. For example, for the AFM images (see section 4.a) where the surface heights are measured on nanometer scale, each bin has a fixed constant width of 1, since we do not intend to capture changes of less than 1 nm.

### 3. Synthetic Surfaces

#### 3.1 Data

Prior to applying the proposed measure at experimentally fabricated nanostructured surfaces, the measure is validated via applications where there is some form of intuition regarding the state of maximum complexity. As mentioned previously, we seek to attribute maximum complexity at states between total order and total randomness. Therefore, the measure is firstly tested on synthesized nanostructured surfaces with controlled mixture of randomness and order to validate its performance.

Three series of synthesized surfaces are used. In the first series, the full order is represented by a homogeneous flat surface, and noise is added by gradually and randomly replacing surfaces points with values taken from a Gaussian noise distribution. In the second case, the full order is represented by an almost homogeneous surface with very large correlation length, and the randomness is introduced through a reduction of correlations (smaller correlation length), resulting in the random morphology of almost zero correlation length. In the last case, we construct mounded periodic surfaces consisting of mounds positioned on the nodes of a regular grid. Randomness is imposed by changing randomly the positions, widths and heights of the mounds on the surface, by adding noise to their values. The gradual increase of noise amplitude degrades periodicity and order and leads to the full randomization of surface morphology.

#### 3.2 Results

##### 3.2.1 Partially noisy surfaces

The initial surface is flat, and its points are gradually replaced in 10 iterations with Gaussian white noise by changing 10% at each iteration. This creates 9 intermediate states between order and randomness. Fig. 2 shows four indicative surfaces with side length L=200 nm, having noise fraction 20%, 50%, 80% and 100% respectively. The bin width is set to 0.1 and the maximum scale is chosen to be $R = 20$.

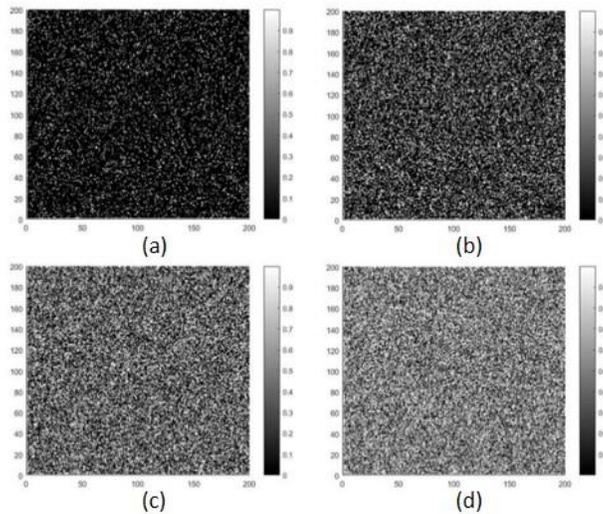

Fig. 1. Synthetic surfaces produced by gradually and randomly replacing the 20% (a), 50% (b), 80% (c) and 100% (d) of the points of the initial 2-dimensional flat surface $z(x,y) = 0$ with Gaussian white noise.

The result of the calculation of the complexity measure versus the fraction of noisy points is shown in Fig. 3. The diagram clearly demonstrates that the higher complexity is assigned to noise fractions about 60-70% between the full order (i.e. flat surface) and the total noisy surface. Therefore, in this example of synthetic surfaces, the proposed metric of complexity meets the requirement of maximization between order and randomness in a state of increased heterogeneity.

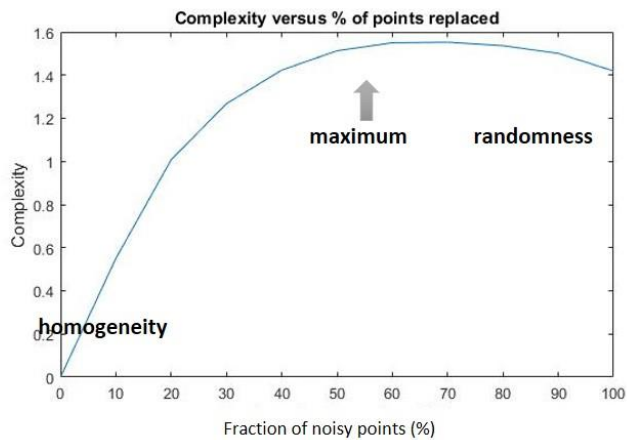

Fig. 2. The complexity measure versus the fraction of noise points on a flat surface. It clearly exhibits its maximum value at surfaces arising after the replacement of 60%-70% of the flat surface points with noise. This example clearly demonstrates the difference between full randomness and complexity.

### 3.2.2 Surfaces with decreased correlation length

The second application used for the validation of the proposed complexity measure refers to the construction of isotropic rough surfaces that have $L^2$ points. The purpose is to generate a surface that is morphologically rich, having correlation and stochasticity at the same time. This is achieved by using Gaussian autocorrelation functions. Random states are approached by gradual decrease of the correlation length which quantifies the range of correlation on surface. The initial rough surface has a side of length $L = 200\ nm$, the bin width is set to 0.1 and the maximum scale is $R = 60\ nm$. The correlation length spans from 2 to 72 $nm$. The transition from randomness to order is depicted in Fig. 4:

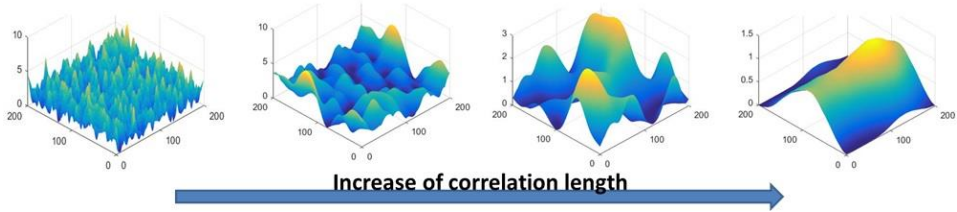

Fig. 3. The left surface has very limited correlations between its points, resulting in a very noisy outcome. As the correlation length is increased, the surfaces are far less noisy, approaching order. It is expected that a surface lying between order and noise has the maximum complexity.

Subsequently, complexity is calculated for different surfaces resulting from the change of the correlation length. For more reliable results, complexity is calculated as the average complexity of 80 surfaces constructed for each correlation length. The results are shown in Fig. 5 where the proposed complexity measure exhibits its maximum at surfaces with intermediate correlation lengths that are neither clearly ordered nor fully random, but present the highest degree of heterogeneity for the specific scales chosen in this study.

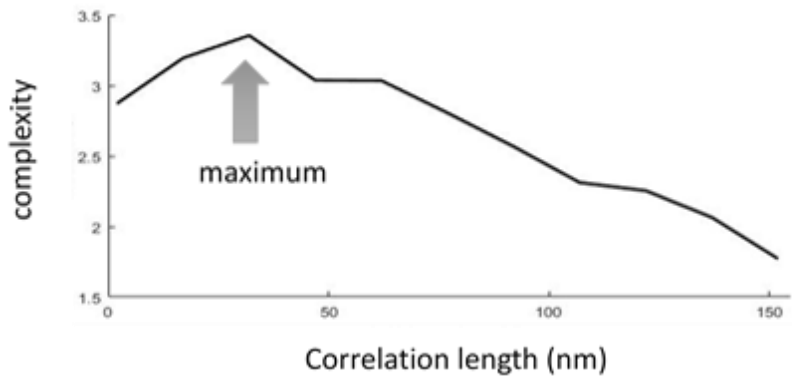

Fig. 4. Complexity versus correlation length for $R = 60$ nm verify that surfaces between order and randomness get higher complexity measure.

### 3.2.3 Mound surfaces with increased positional and shape randomness

A mounded surface is generated via the successive addition of Gaussian peaks on a flat surface which are placed at the nodes of 2D square grid. For each peak, three parameters are considered:

- The peak position
- The height of surface peaks
- The width of the peak

The fully ordered (isotropic) surface has a side of length $L = 500\ nm$, 81 gaussian peaks of height $h = 3\ nm$, and a width of $w = 10\ nm$. This is shown in Fig. 6 (a).

By gradually randomising the positions, widths, and heights of the peaks (surfaces (b) and (c)), a transition to a fully random surface can be achieved (surface (d)) (see Fig. 6).

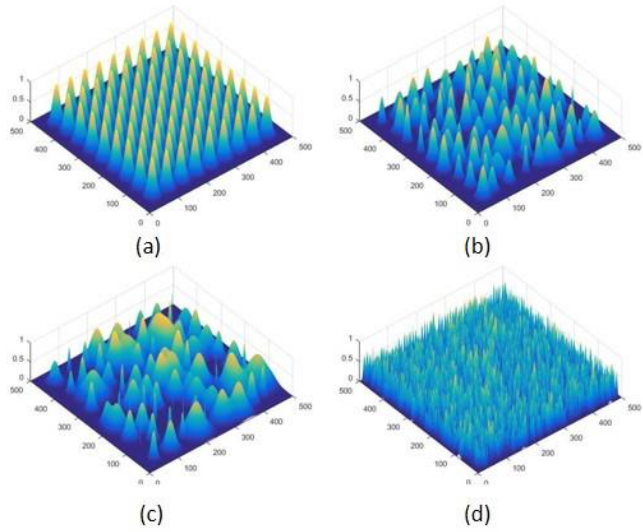

Fig. 5. Initial fully periodic (ordered) mound surface (a) and the surfaces that are generated after imposing randomness on their spatial (position) and shape (width and height) parameters (b-d).

The complexity measure calculated by the modified Alamino's method for these surfaces is shown in Figure 7. It was found that it exhibits a maximum between order and randomness justifying its performance in these model conditions.

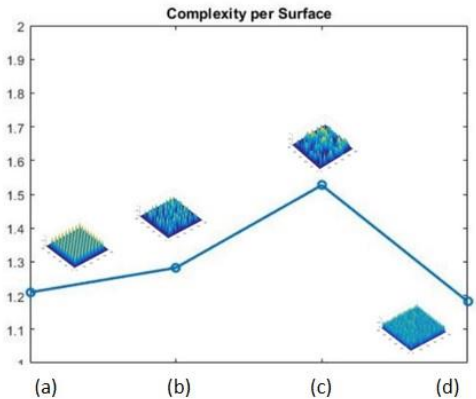

Fig. 6. Complexity measure calculated for mounded surfaces with increased randomness of mound positions and shapes. Notice the maximum of complexity lying between the two extrema of full order and randomness.

4. **Experimental Surfaces**

   4.1 **Polymer surfaces after plasma etching**

      4.1.1 **Data**

Besides the aforementioned synthesized surfaces, the complexity measure has additionally been calculated using 'real' laboratory fabricated surfaces of Poly(Methylmethacrylate) (PMMA). PMMA is a polymeric material, and for this fabrication it was etched $O_2$ in a dry plasma reactor with 100 sccm $O_2$, 0 Bias Voltage, 2000W generator power, 0.75 Pa Pressure. These samples were measured and inspected with AFM, using a CP-II Veeco in a non-contact tapping mode measurement. The polymer morphology exhibits an increased complexity with etching time. This complexity cannot be sufficiently described using conventional mathematical tools. A more detailed description of the nanostructured polymer surfaces and the corresponding analysis using Fourier transform, was published in previous work of the group (see [12]). The AFM measurements were acquired on a square grid of 512x512 points with side equal to 2μm (with the only exception being the measurement at 10min, where due to the size of the structures the side was chosen to be 4μm). The upper scale in entropy analysis was kept fixed at $R = 200\ nm$. These surfaces are shown in Fig. 8.

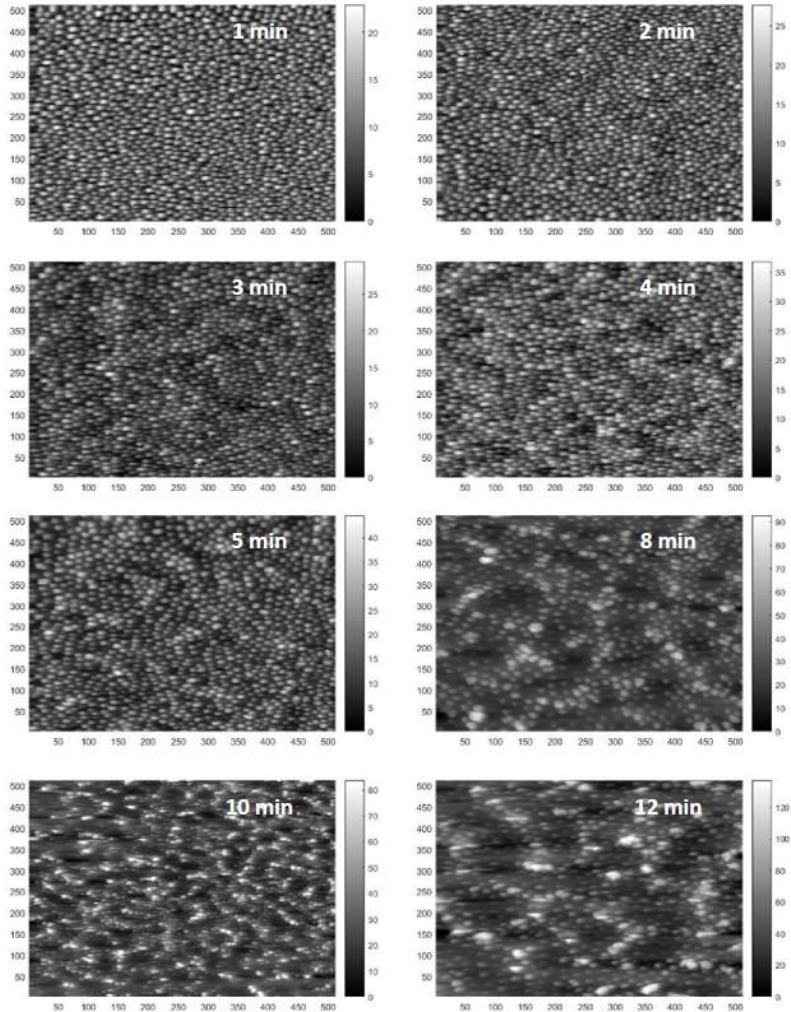

Fig. 7: AFM images of PMMA surfaces for different times of $O_2$ plasma etching.

### 4.1.2 Results

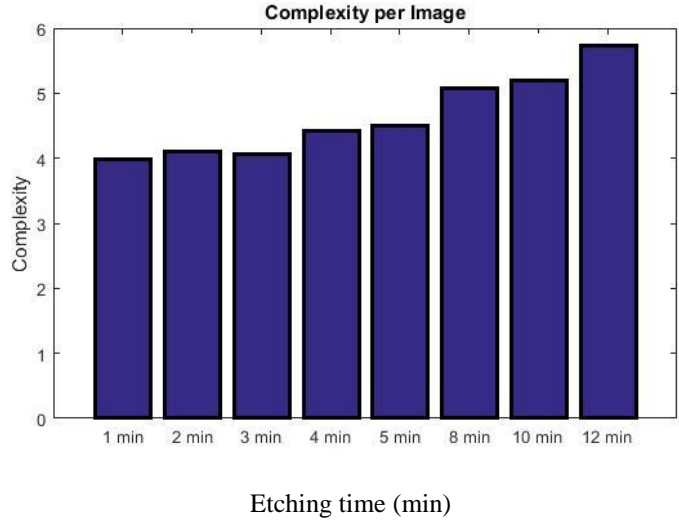

Etching time (min)

Fig. 8. Complexity measure of PMMA surface nanomorphologies versus plasma etching time. The AFM measurements were a 2x2 μm square grid of 512x512 points (the only exception is the measurement at 10min where the side is 4μmx4μm). As expected, it increases monotonically with etching times, while it seems that it categorizes surfaces in three clusters (1-3min, 4-5min and 8-12min) each of which shows a different complexity behavior.

The calculated complexity for the PMMA surfaces versus etching time is shown in Fig. 9. Not surprisingly, it exhibits an increasing trend in agreement with the behavior of other roughness parameters such as rms, skewness, kurtosis [12]. Besides the increasing trend, one can also notice the clustering of complexity values in three groups for etching times (1-3)min, (4-5)min and (8-12)min. To better understand this clustering, we should describe in more detail the evolution of surface morphology with etching time. At short times, we have a well-ordered arrangement of nanopillars on the polymer surface with similar heights. This regular pattern is distorted at 4min from the gradual appearance of larger pillars that start to form a second hierarchical level in surface topography. At etching times of around 8min, the high pillars start to form a network with their aggregates. This network changes the features of the second hierarchical level, and this change can be clearly detected and captured in the behavior of the complexity measure as shown in Fig.9.

### 4.2 Micro- and nanostructured Aluminium surfaces

#### 4.2.1 Data

The complexity measure has also been applied to surfaces of Aluminium, which are etched with a solution of $FeCl_3$ [13] or $CuCl_2$. Aluminium (> 99.5% Al) substrates are first cleaned with standard procedures, followed by immersion into the corresponding etchant of 1 M concentration with ultrasonication. The samples are subsequently cleaned to remove residue and unreacted species. Immersion into boiling water further creates the nanostructures. The surfaces coming from the etching of two solutions are inspected with top-down SEM images and are about $L = 54\ \mu m$ (see Figures 10 and 11 respectively). The maximum scale used in the complexity estimation has been chosen to be $R = 1.7\ \mu m$.

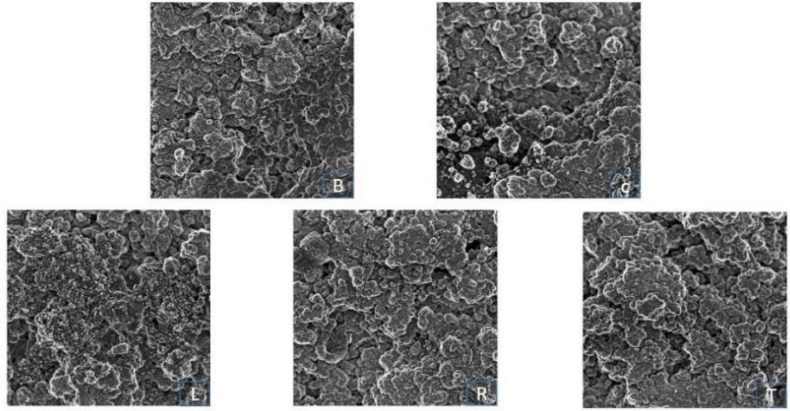

Fig. 9. SEM images from different areas on the surface of an Al sample when CuCl$_2$ solution is used for etching: B stands for bottom, C for center, L for left, R for right and T for top.

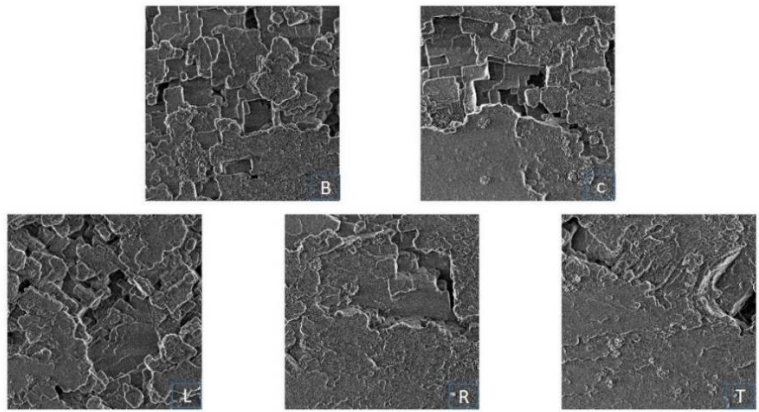

Fig. 10. SEM images from different areas on the surface of a Al sample when FeCl$_3$ solution is used for etching: B stands for bottom, C for center, L for left, R for right and T for top

### 4.2.2 Results

The average values of the complexity measure that are calculated for the sets of SEM images etched with CuCl$_2$ and FeCl$_3$ (Figures 10 and 11 respectively) are shown in Figure 12. It seems that the results validate the expected increase in complexity of the Al surfaces when etched with CuCl2 over those etched with FeCl3.

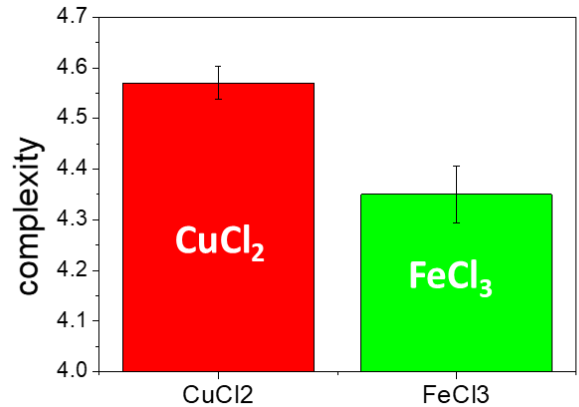

Fig. 12: The average complexity of the two sets of SEM images shown in Figures 11 and 12 for $CuCl_2$ (red bar) and $FeCl_3$ (green bar) etching solutions respectively. According to the results, the Al surfaces that are generated by etching with $CuCl_2$ are clearly more complex than those that are produced by the use of $FeCl_3$.

The increased complexity and heterogeneity of Al surfaces after their wet etching with the $CuCl_2$ canalso have an impact on their properties. This was validated when we measured their condensation heat transfer coefficient and found that the $CuCl_2$ surfaces exhibited clearly higher heat transfer coefficients at all subcooling values (see Fig. 13), defined as the difference between the steam temperature and the surface temperature. This is in line with the measured complexities, and evidences the link between the spatial complexity of micro and nanostructured Al surfaces and their heat transfer coefficient. It is important here to mention that these surfaces were hydrophobized with an ultrathin (≈40 nm) hydrophobic coating to promote jumping dropwise condensation. Such a thin coating is expected to induce minimal change in the surface roughness. To achieve this an initiated chemical vapor deposition method was used. The surfaces were then subjected to low-pressure saturated steam and condensation was initiated by cooling the samples.

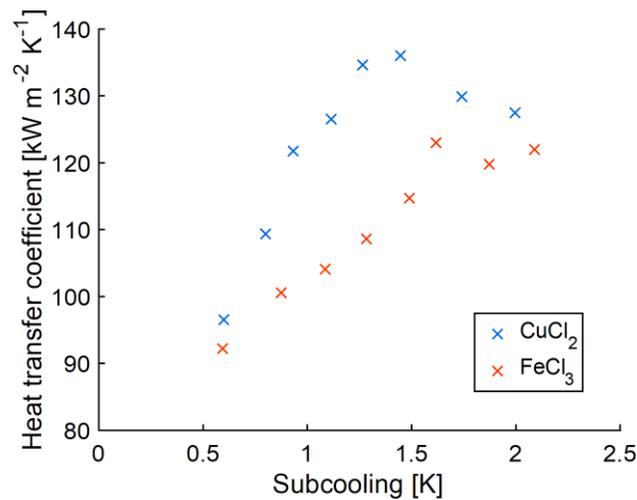

Fig. 13: The heat transfer coefficient of Al surfaces etched with $CuCl_2$ (blue crosses) and $FeCl_3$ (red crosses) measured for different subcooling temperatures. The surfaces etched by $CuCl_2$ exhibit larger heat transfer coefficient in agreement with their higher complexity and morphological heterogeneity shown in Fig.13.

## 5. Conclusions

In this work, inspired by recent advances in the complexity science, we proposed a measure for the complexity of nanostructured surfaces based on the concept of deviation from the average/statistical symmetry. We presented the methodology for its estimation, using the multiscale entropy of their microscopy images. Subsequently, in order to validate the method's ability to distinguish between complexity and randomness, and maximize on surfaces with more heterogeneous combination of order and randomness, we applied the proposed methodology and calculated the complexity measure in a series of synthesized morphologies with controlled contribution of randomness. Finally, we investigated real experimental surfaces of PMMA and Al after their micro and nanostructuring processing, and demonstrated the potential of our complexity approach in quantifying the impact of processing conditions on their morphologies, as well as its ability to explore their link to specific surface properties. Advanced functionalities of micro and nanostructured surfaces are strongly linked to the geometric characteristics of their morphologies. The latter usually exhibit complicated structures consisting of both ordered and random components. The quantitative characterization of this type of structures is a necessary step in order to achieve their controlled fabrication and targeted applications. With our method this desired quantitative characterisation can be materialized. This can lead to a better understanding of the surface characteristics and the topography of complex structures not only in surfaces of nanostructured materials, but can very easily be translated and applied in a plethora of applications in numerous disciplines.

We plan to continue this work in two directions. The first is to deepen the theoretical foundation of nanocomplexity putting more emphasis on the role of scale in complexity definition and estimation. The second direction concerns more applications in experimental surfaces focusing on the link of surface complexity with the processing steps and functional properties.

**Acknowledgement**

This work has received funding from the European Union's Horizon 2020 research and innovation programme under grant agreement No 801229 (project Harmonic: HierARchical Multiscale NanoInterfaces for enhanced Condensation processes)